\documentclass[
 amsmath, amssymb, breqn,
 aps, prl, superscriptaddress, twocolumn,
longbibliography]{revtex4-1}

\usepackage{graphicx}% Include figure files
\usepackage{dcolumn}% Align table columns on decimal point
\usepackage{bm}% bold math
\usepackage{upgreek}
\usepackage{braket}

\begin{document}

\title{Elimination of Thermomechanical Noise in Piezoelectric Optomechanical Crystals}

\author{H. Ramp}
\email{ramp@ualberta.ca}
 \affiliation{Department of Physics, University of Alberta, Canada} 
 
\author{B.D. Hauer}%
 \affiliation{Department of Physics, University of Alberta, Canada}

\author{K.C. Balram}
\affiliation{Center for Nanoscale Science and Technology, National Institute for Standards and Technology, Gaithersburg, USA}

\author{T.J. Clark}%
 \affiliation{Department of Physics, University of Alberta, Canada}

\author{K. Srinivasan}
\affiliation{Center for Nanoscale Science and Technology, National Institute for Standards and Technology, Gaithersburg, USA}

\author{J.P. Davis}
\email{jdavis@ualberta.ca}
 \affiliation{Department of Physics, University of Alberta, Canada} 

\date{\today}

\begin{abstract}
Mechanical modes are a potentially useful resource for quantum information applications, such as quantum-level wavelength transducers, due to their ability to interact with electromagnetic radiation across the spectrum.  A significant challenge for wavelength transducers is thermomechanical noise in the mechanical mode, which pollutes the transduced signal with thermal states. In this paper, we eliminate thermomechanical noise in the GHz-frequency mechanical breathing mode of a piezoelectric optomechanical crystal using cryogenic cooling in a dilution refrigerator. We optically measure an average thermal occupancy of the mechanical mode of only $0.7\pm0.4 ~ \mathrm{phonons}$, providing a path towards low-noise microwave-to-optical conversion in the quantum regime.

\end{abstract}

\pacs{Valid PACS appear here}% PACS, the Physics and Astronomy
                             % Classification Scheme.
\keywords{Suggested keywords}%Use showkeys class option if keyword
                              %display desired
\maketitle

Quantum information science may have begun with atoms and optical photons, but it has expanded to include numerous experimental platforms that have demonstrated quantum behavior.  Examples of quantum devices now include a wide array of well-controlled natural systems including neutral atoms \cite{Saffman2016}, ions \cite{Blatt2008,Monroe2013}, electronic \cite{Epstein2005,Hanson2006} and nuclear spins~\cite{Kane1998,Childress2006}, as well as fabricated systems such as quantum dots \cite{Okazaki2016}, superconducting circuits \cite{Clarke2008,Schoelkopf2008,Martinis2009}, and mechanical devices \cite{OConnell2010,Wollman2015,Lecocq2015,Reed2017}. Each system has a unique set of properties that make them advantageous for specific applications: for example, long coherence times make atomic systems a natural candidate for quantum memories \cite{Julsgaard2004, Lvovsky2009, Simon2010, Saglamyurek2018}, while the flexible nature of fabrication makes superconducting circuitry ideal for creating quantum processing gates \cite{DiCarlo2009}. This has culminated in the vision of hybrid quantum systems that can link multiple sub-systems into complex quantum machines \cite{Kimble2008,Kurizki2015}.

\addtolength{\textfloatsep}{-0.1in}
 
One major challenge in creating hybrid quantum systems arises from transferring quantum information between the different sub-systems.  Photons are the obvious medium for transferring of information as most quantum systems can interact with light, however the relevant wavelength varies widely.  Furthermore, quantum information has been effectively transferred over long distances using optical photons \cite{Ursin2007,Valivarthi2016,Yin2017}.  This has spurred interest in the development of mechanical wavelength transducers -- devices that use photon-phonon interactions to coherently convert photons between different wavelengths while preserving quantum information. In particular, interest lies in converting between the microwave frequencies used in superconducting quantum processors and telecom-wavelength optical photons used in fiber networks for the purposes of networking quantum processors \cite{Kimble2008}.  At its core, a mechanical wavelength transducer consists of a mechanical element coupled to two electromagnetic resonances at the desired input/output wavelengths. State-of-the-art transducers have demonstrated wavelength conversion for classical signals within the optical wavelength range~\cite{Hill2012, Dong2012, Liu2013}, within the microwave regime~\cite{Lecocq2016,Ockeloen-Korppi2016}, the up-conversion of microwave tones to optical wavelengths~\cite{Bochmann2013,Bagci2014, Balram2016}, and the bidirectional conversion of microwave and optical signals~\cite{Andrews2014, Higginbotham2018, Vainsencher2016}.  However, no system has demonstrated the characteristics needed to perform microwave-to-optical conversion in the quantum regime without added thermomechanical noise.  

\begin{figure}[b]
    \includegraphics[width=.47\textwidth]{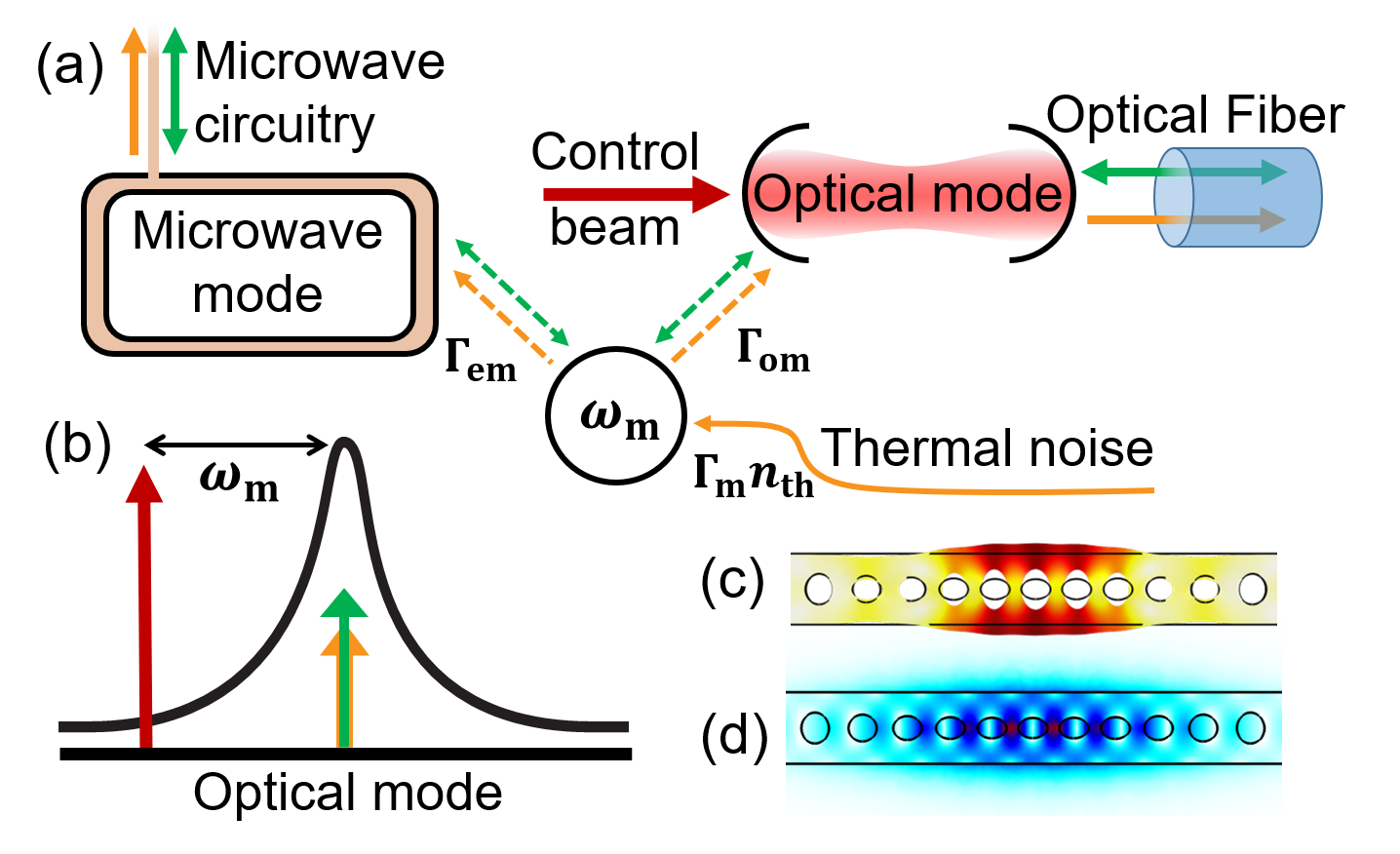}
    \caption{(a) Schematic of a route to mechanically-mediated wavelength conversion: a mechanical resonator at frequency $\omega_\mathrm{m}$ is coupled to microwave and optical modes (for example) at rates $\Gamma_\mathrm{em}$ and $\Gamma_\mathrm{om}$ respectively, allowing for the bidirectional transduction of a quantum state (green). The mechanical resonator is populated with thermal phonons (orange) at a rate $\Gamma_\mathrm{m}n_\mathrm{th}$, which are converted into thermomechanical noise photons. (b) A control beam (red) is red-detuned from the optical resonance by one mechanical frequency, allowing conversion between phonons in the mechanical mode and optical photons. (c) Displacement simulation of the GHz-frequency mechanical breathing mode and (d) electric field simulation of the optical mode of an optomechanical crystal.}
    \label{fig:Device}
    \centering
\end{figure}

Transitioning from the conversion of classical tones to the quantum conversion of single photons is daunting, in part due to the probability of extracting thermal phonons from the mechanical mode instead of the desired quantum state, as outlined in Fig.~1. For this reason, it is important to eliminate thermomechanical noise such that the mechanical resonator is in its ground state, where the number of thermal phonons is less than one.  Compatibility between passive cooling into the ground state and measurement with optical wavelengths has proven difficult, in large part due to heating from optical absorption \cite{Meenehan2014,Meenehan2015}. Here, we demonstrate the ground state cooling of a GHz-frequency mechanical mode of a GaAs optomechanical crystal \cite{Balram2014}, as measured using telecom wavelength light, reaching an average thermal phonon occupation of $\braket{n} = 0.7\pm0.4$. With the additional fact that GaAs is piezoelectric and can be interfaced with microwave signals, this telecom wavelength measurement of a mechanical resonator in its quantum ground state opens up the possibility of low-noise microwave-to-optical conversion in the quantum regime. 

\begin{figure}[t]
    \centering
    \includegraphics[width=0.45\textwidth]{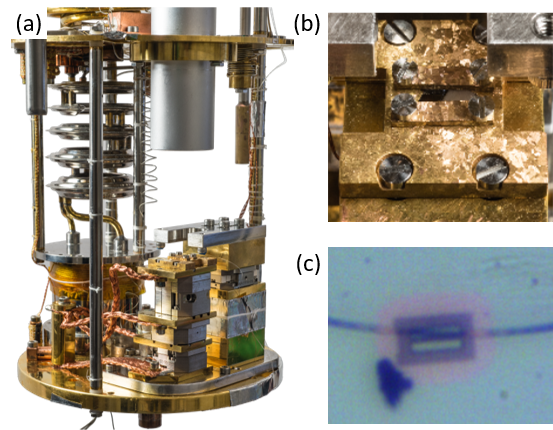}
    \caption{(a) Base plate of the dilution refrigerator. Devices are mounted in a chip holder, placed on a stack of piezoelectric positioning stages. Copper braids provide thermal coupling between the chip holder and the dilution unit. A high-efficiency dimpled tapered fiber is mounted next to the positioning stack. An optical imaging system~\cite{MacDonald2015} above the chip holder allows for real-time optical access to the device to facilitate coupling. (b) Close up of holder with GaAs chip. c) Microscope image of device while coupled to a dimpled tapered fiber.   }
    \label{fig:measSetup}
\end{figure}

The elimination of thermomechanical noise in the 2.4 GHz breathing mode of the GaAs optomechanical crystal is achieved using a carefully designed refrigeration system (shown in Fig.~2 and detailed in Appendix A) to best ensure that the GaAs chip is thermalized to the base plate of the dilution refrigerator.  Despite this precaution, low-power optical measurements of the $1550~\mathrm{nm}$ optical mode of the device result in heating of the mechanical mode due to optical absorption, which prevents access to the ground state during continuous wave measurements, as has been previously observed \cite{Meenehan2015,Hauer2018}.  To mitigate this, we perform pulsed measurements to measure the mechanical motion in the form of the time-dependent integrated power spectral density \cite{Hauer2015},
\begin{equation}
    \label{eqn:calibration}
    \int S_V(\omega,t)d\omega = \alpha \braket{n}(t)+\beta,
\end{equation}
which is linearly related to the (time-dependent) average phonon occupancy of the mechanical mode, $\braket{n}(t)$.  Here $\alpha$ is a conversion factor determined by device mechanical properties and the measurement setup, and $\beta=(S_{V}^\mathrm{imp}+S_V^\mathrm{gs} + S_V^\mathrm{ba})\Delta\omega$ represents noise due to temperature-independent measurement imprecision $S_{V}^\mathrm{imp}$, ground state motion $S_V^\mathrm{gs}$, and measurement backaction $S_{V}^\mathrm{ba}$, integrated over the measurement bandwidth $\Delta\omega$ \cite{Hauer2018}. The measurement imprecision fundamentally arises from shot noise on the mechanical mechanical mode \cite{Jacobs2014} and is further increased by internal optical cavity loss and imperfect heterodyne detection \cite{Teufel2016}.

\begin{figure}[b]
    \includegraphics[width=.45\textwidth]{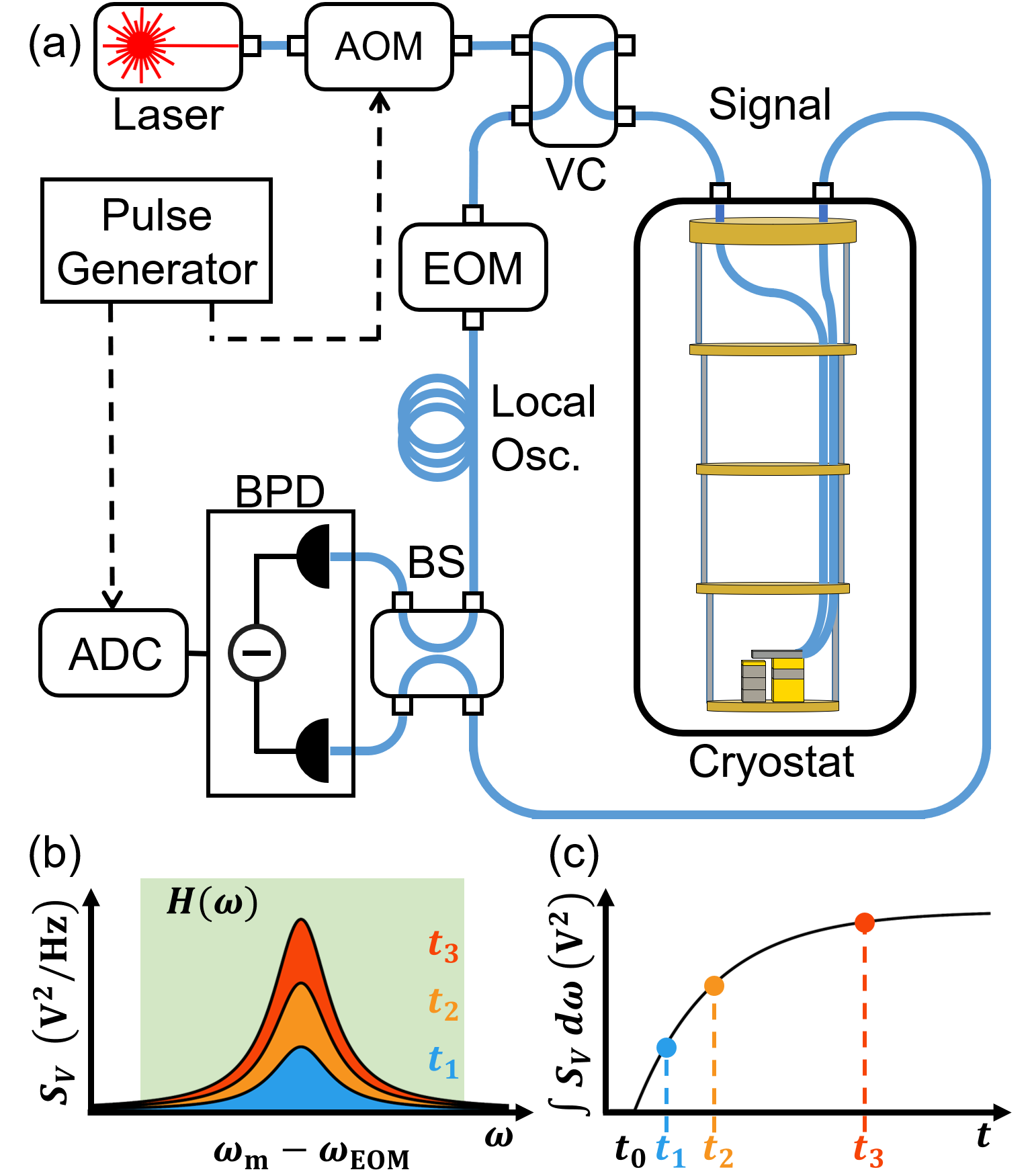}
    \caption{(a) Simplified schematic of optical heterodyne detection. AOM: acousto-optic modulator, VC: variable coupler, EOM: electro-optic modulator, BS: beamsplitter, BPD: balanced photodiode, ADC: analog-digital converter. (b) Cartoon of frequency-domain mechanical signal at three different times during a pulse. The mechanical signal is convolved with the green filter $H(\omega)$ to obtain (c) the time-dependent mechanical area.}
    \label{fig:Optical_setup}
    \centering
\end{figure}

\begin{figure*}[t]
    \centering
    \includegraphics[width=.975\textwidth]{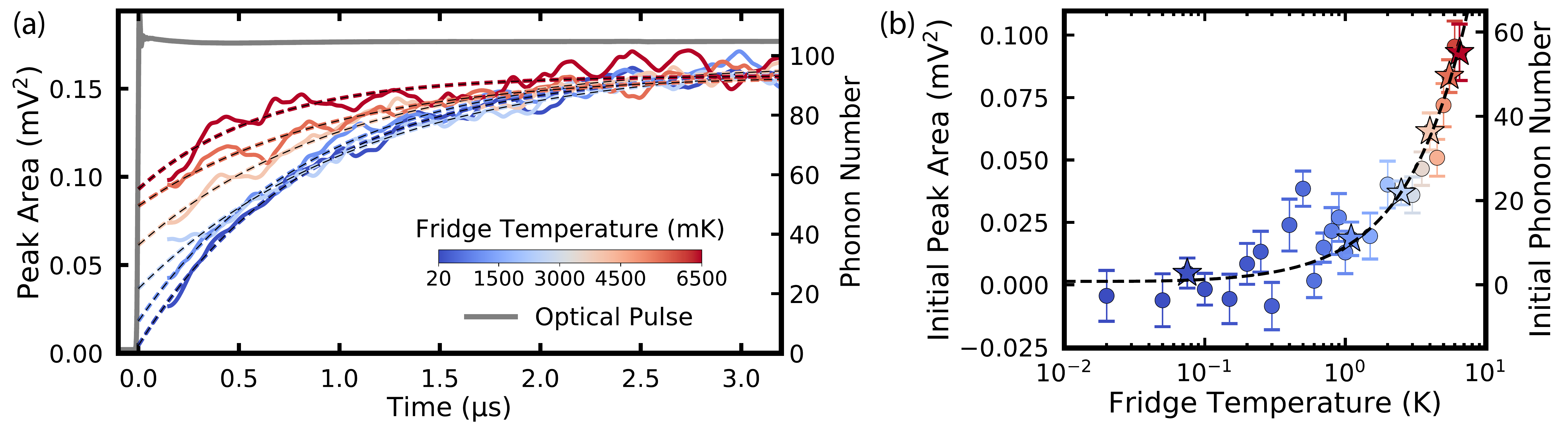}
    \caption{(a) Heterodyne pulsed measurements showing the thermomechanical noise of the mechanical mode as a function of time. Measurements are truncated to begin at $0.25~\upmu\mathrm{s}$ after the optical pulse (grey) due to the $6.25~\mathrm{MHz}$ bandwidth of the filter function. Fits to Eqn.~\eqref{eqn:phonongrowth} (dashed lines) are used to extrapolate the pulses back to $t=0$. Left axis presents the data in terms of power spectral density peak area. Right axis shows recalibrated data in terms of phonon number. (b) The peak areas at the onset of the optical pulse, color-coded to match the fridge temperature scale, with stars to denote the example traces from (a). Peak areas from high temperature data $T \ge 1.5~\mathrm{K}$ are used to calibrate the initial peak area to phonon number. The complete data set is fit to the Bose-Einstein distribution with an offset (black dashed) to determine the average number of phonons in the mechanical mode at $20~\mathrm{mK}$.   }
    \label{fig:Pulsedmechanics}
\end{figure*}

The pulsed measurement is performed by setting the measurement laser to the optical resonance frequency $\omega_\mathrm{c}$, which is split into two equal-length paths by a variable coupler (Fig.~3a).  The signal in the local oscillator arm is modulated by an electro-optic modulator (EOM) driven near $V_\pi$ to create sidebands at $\omega_\mathrm{c}\pm\omega_\mathrm{EOM}$.  The signal in the device arm interacts with the optomechanical resonator to create sidebands at $\omega_\mathrm{c}\pm\omega_\mathrm{m}$. When mixed at the beamsplitter, the arms beat together to create a signal at $\omega_\mathrm{m}-\omega_\mathrm{EOM}$, which we typically set to $30~\mathrm{MHz}$ to allow detection using a low-frequency balanced photodiode. The photodiode voltage signal is then measured in the time domain with a fast analog-digital converter. This heterodyne measurement is phase-sensitive, where drifts in path length between the local oscillator and signal arms cause changes in measurement amplitude. To circumvent this for continuous measurements, the path lengths, and therefore relative phase, can be locked using a fiber stretcher in the local oscillator arm, which is controlled using a feedback loop that locks to the error signal of the balanced photodiode. For fast pulses this locking technique is not used because it relies on continuous optical input to the signal arm.  Nonetheless, the phase does not drift on the short time scales of a single pulsed measurement. To account for longer term drifts, each measurement is comprised of 20,000 averaged repetitions.

During each optical measurement pulse, the number of phonons in the mechanical mode transitions from an initial phonon number -- set by coupling to the ambient thermal bath of the dilution refrigerator -- to a new thermal equilibrium set by the optical absorption heating and the cooling power of the system. The time dependence of this heating can be measured through the time-dependent integrated power spectral density, Eqn.~(1).  Fig.~\ref{fig:Optical_setup}b shows a cartoon of the mechanical power spectral density at three different times during an optical pulse. Behind the spectra, $H(\omega)$ shows the $6.25~\mathrm{MHz}$ bandpass filter that is convolved with the mechanical signal to obtain the mechanical peak area as a function of time. 

Following the work of Refs.~39 and 41, the average number of thermal phonons in the mechanical mode $\braket{n}(t)$ can be modelled by considering coupling to the ambient fridge bath with $n_\mathrm{th}$ phonons at rate $\Gamma_\mathrm{m}$, and a laser-induced hot phonon bath of $n_\mathrm{p}$ phonons coupled to the mechanics at a rate of $\Gamma_\mathrm{p}$. The simplest model to which one could fit this heating is an instantaneous onset of the hot phonon bath at the beginning of the optical pulse (at time $t_0$).  The mechanical mode occupancy $\braket{n}(t)$ then increases from its initial occupancy $\braket{n}(t_0)$ to the thermal equilibrium $n_\mathrm{eq} = (\Gamma_\mathrm{m} n_\mathrm{th} + \Gamma_\mathrm{p} n_\mathrm{p})/{\Gamma}$ at a rate $\Gamma = \Gamma_\mathrm{m} + \Gamma_\mathrm{p}$ according to,
\begin{equation}
    \label{eqn:phonongrowth}
    \braket{n}(t) = \braket{n}(t_0)e^{-\Gamma(t-t_0)}+n_\mathrm{eq}(1-e^{-\Gamma(t-t_0)}).
\end{equation}
In Fig.~\ref{fig:Optical_setup}c, a cartoon of the mechanical peak area $\int S_{V}(\omega,t)~d\omega$ is shown according to Eqn.~\eqref{eqn:phonongrowth} over the duration of an optical pulse to demonstrate the peak area growth associated with optical heating of the mechanical mode. By varying the temperature of the thermal reservoir -- here the base-plate of the dilution refrigerator -- we can track not only the time-dependent phonon number $\braket{n}(t)$ but also the initial thermal phonon occupation $\braket{n}(t_0)$.

\addtolength{\textfloatsep}{0.1in}

We present such pulsed heterodyne measurements of the GaAs optomechanical crystal in Fig.~\ref{fig:Pulsedmechanics}a, at dilution refrigerator temperatures between $20~\mathrm{mK}$ and $6.5~\mathrm{K}$. In these measurements, a $1.5~\upmu\mathrm{W}$ optical laser pulse (set by the minimum power needed for adequate signal-to-noise) is turned on at $t=t_0$ and populates the optical mode with $n_\mathrm{cav}\approx230$ photons on a timescale of $1/\kappa = 32~\mathrm{ps}$.  Absorption of photons in the optical mode causes the device to consistently heat to a temperature above $6.5~\mathrm{K}$ regardless of the initial starting temperature. Each trace of the peak area is fit to Eqn.~\eqref{eqn:phonongrowth} to extract the initial and final mechanical peak areas, proportional to the phonon occupancy.

At high temperatures, we assume that the device is thermalized to the dilution refrigerator when the measurement begins, which is supported by the linear relationship in Fig.~4b. However, we do not assume that the device is thermalized at millikelvin temperatures, where the GaAs thermal conductivity drops significantly due to $T^3$ scaling~\cite{Blakemore1982}. For this reason only temperatures $T\ge1.5~\mathrm{K}$ are used in the phonon number calibration. We fit the initial peak areas of the high temperature data using Eqn.~\eqref{eqn:calibration} to determine the conversion factor $\alpha$ as well as the total noise $\beta = 1.18~\mathrm{mV}^2$. A second measurement, away from the mechanical resonance, is used to find $S_V^\mathrm{imp}$ in the absence of $S_V^\mathrm{gs}$ and $S_V^\mathrm{ba}$. From this we find ${S_V^\mathrm{imp}}/{(S_V^\mathrm{ba}+S_V^\mathrm{gs})} = 1.6$, which implies these measurements are performed near the standard quantum limit.

In Fig.~\ref{fig:Pulsedmechanics}b, the initial peak areas are plotted with $\beta$ subtracted (left axis) and recalibrated to initial phonon number (right axis) using the conversion factor $\alpha$. The initial phonon numbers are then fit using a Bose-Einstein distribution to find the thermal offset between the device and the fridge at base temperature. At a fridge temperature of $20~\mathrm{mK}$ we find that the mean phonon occupancy is initially $\braket{n}(t_0) = 0.7\pm0.4$, with $95\%$ confidence from the fit. This implies that the mechanical mode is in the ground state $59\%$ of the time and that the device thermalizes to $0.13\pm0.05~\mathrm{K}$ when the fridge thermometry reads $0.02~\mathrm{K}$. Applying the calibration to the time-resolved measurements in Fig.~\ref{fig:Pulsedmechanics}a shows that the mechanical mode saturates to a thermal occupancy of 95 phonons in $3~\upmu\mathrm{s}$. 

Our measurements demonstrate that the intrinsic properties of the GaAs optomechanical crystal show promise for the application of low-noise quantum wavelength transduction. The intrinsic cooperativity $\mathcal{C} = \Gamma_\mathrm{om}/{\Gamma_\mathrm{m}}$ \cite{Aspelmeyer2014} describes the relative rates of information transfer between mechanics and optics $\Gamma_\mathrm{om}/{2\pi} = 0.31~\mathrm{MHz}$ to the mechanical damping $\Gamma_\mathrm{m}/{2\pi}=83~\mathrm{kHz}$ (Appendix B). During our measurements we find $\mathcal{C} = 3.7$. An intrinsic cooperativity exceeding $1$ suggests that in the absence of the laser-induced hot phonon bath, GaAs optomechanical crystals are a leading candidate for high-efficiency, low-noise wavelength conversion. However, for the coherent transfer of quantum states, the relevant figure of merit must include the total decoherence rate $\Gamma n_\mathrm{eq}$. We include the total damping rate $\Gamma = 1.05~\mathrm{MHz}$ and equilibrium phonon number $n_\mathrm{eq}=95$ to calculate the quantum cooperativity $\mathcal{C}_\mathrm{qu} =  \Gamma_\mathrm{om}/{\Gamma n_\mathrm{eq}}$ to be $ 3\times10^{-3}$. 

From these cooperativities, we calculate the added thermomechanical noise~\cite{Hill2012,Andrews2014} from the ambient dilution refrigerator bath $n_\mathrm{add,th} \approx 0.4 ~\mathrm{quanta}$ and from the combined laser-induced and ambient baths $n_\mathrm{add} \approx 262 ~\mathrm{quanta}$ (Appendix B). The requirement of high cooperativity $\mathcal{C}_\mathrm{qu}>1$ for low-noise is illustrated by comparing the number of added noise quanta to the number of phonons in the respective baths: high intrinsic cooperativity diminishes the 0.7 phonon bath to 0.4 added quanta, whereas low quantum cooperativity increases the effect of the 95 phonon bath to 262 added quanta. 

The use of optical pulses to determine the phonon occupancy of the GaAs optomechanical crystals reveals several limitations that require further investigation. First, the intrinsic mechanical damping rate was expected to be lower than the measured value of $\Gamma_\mathrm{m}/2\pi=83~\mathrm{kHz}$ at millikelvin temperatures. The high damping rate could result from any number of sources, including surface roughness \cite{Balram2014}, two-level systems \cite{Hamoumi2018}, or clamping losses \cite{Nguyen2013}.  Further studies would be required to elucidate.  Second, we have identified the laser-induced hot phonon bath as a limiting factor for microwave to optical transduction. Techniques such as passivating the GaAs surface to reduce roughness and the influence of mid-gap surface states \cite{Guha2017} may result in reduced optical absorption at the surfaces of the optomechanical crystal~\cite{Michael2007}. Limiting optical absorption will in turn reduce the occupancy of the laser-induced bath and may allow GaAs optomechanical crystals to achieve $n_\mathrm{add}<1$ and $\mathcal{C}_\mathrm{qu}>1$ as is required for steady-state low-noise wavelength conversion.

In conclusion, we have used pulsed-optomechanical heterodyne measurements to show that GaAs optomechanical crystals are capable of being cooled to a ground state thermal population of $\braket{n} = 0.7\pm0.4$ phonons using a dilution refrigerator. Optical absorption causes heating of the mechanical mode resulting in a population of $\braket{n} = 95$ thermal phonons in the continuous-wave limit. Despite this heating, the amount of added noise in a potential wavelength conversion application is much less than quantum wavelength transducers with MHz-frequency mechanical modes \cite{Andrews2014}. These results demonstrate GaAs optomechanical crystals are a promising path towards efficient quantum state conversion between microwave and optical light.  We note that a similar conclusion is reached in a simultaneous measurement of GaAs optomechanical crystals \cite{Forsch2018}.

\section*{Acknowledgments}
This work was supported by the University of Alberta, Faculty of Science; the Natural Sciences and Engineering Research Council, Canada (Grants Nos. RGPIN-04523-16, DAS-492947-16, and CREATE-495446-17); Quantum Alberta; and the Canada Foundation for Innovation. B.D.H. acknowledges support from the Killam Trusts.

\section*{Appendix A: Low-temperature setup}

Our low temperature optomechanics setup is designed to allow for flexible optical coupling while also maximizing the thermal connection between the dilution refrigerator and the device. Fig.~\ref{fig:measSetup}a is a photograph of the base plate of the dilution refrigerator which demonstrates the optical coupling and cooling systems. A closeup of the chip holder, Fig.~\ref{fig:measSetup}b, shows the GaAs chip mounted in an annealed copper chip holder that has been gold plated to provide a malleable surface. The chip is screwed in tightly enough to deform the surface of the chip holder, which creates a high surface area mechanical connection for thermal conduction at millikelvin temperatures. The chip holder is connected to flexible copper braids that transfer heat to the base plate of the dilution refrigerator.

The flexibility of the copper braid anchoring system allows the GaAs chip to be freely maneuvered using a 3-axis piezoelectric positioning stack while remaining thermally anchored. This is critical for the dimpled tapered fiber coupling system that allows for optical coupling to any device on the GaAs chip. The optical coupling can be controlled by using the piezoelectric positioning stages to move the device such that the fiber touches different sections of the photonic crystal. Fig.~\ref{fig:measSetup}c shows a photonic crystal nanobeam that is optically coupled to a tapered fiber.

\section*{Appendix B: Optomechanical Characterization}
\label{sec:device_char}

\begin{figure}[t]
    \includegraphics[width=.45\textwidth]{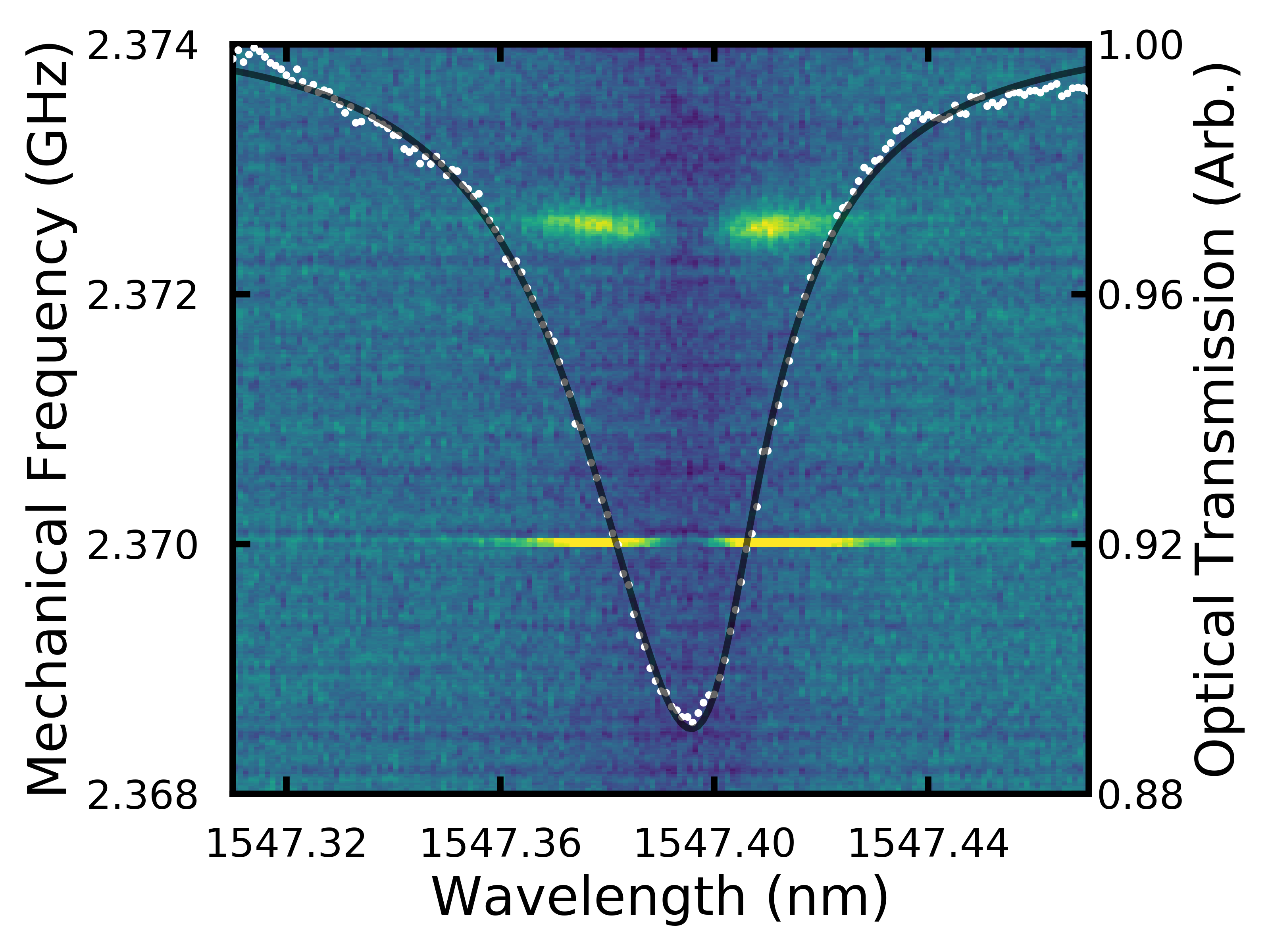}
    \caption{Wavelength scan of the photonic crystal optical mode (white), fit to a Lorentzian curve (black). The background shows the frequency spectrum at each step of the wavelength scan, with the mechanical resonance and EOM calibration tone. }
    \label{fig:Device_optomechanics}
    \centering
\end{figure}

Initial characterization of the GaAs photonic crystal nanobeam was performed at $4.2~\mathrm{K}$ using direct detection. In Fig.~\ref{fig:Device_optomechanics}, a tunable telecom laser was used to probe the photonic crystal optical resonance. A Lorentzian fit was used to extract the center frequency $\omega_c/2\pi = 193.7~\mathrm{THz}$ and the cavity linewidth $\kappa/2\pi = 5.0~\mathrm{GHz}$. The optical scan was performed in discrete steps of $1~\mathrm{pm}$ so that the mechanical spectrum could be measured at every laser detuning. The frequency spectrum has two identifiable peaks: the mechanical mode with $Q_m = 1.1\times10^3$ at frequency $\omega_m/2\pi = 2.3725~\mathrm{GHz}$, and a calibration tone generated by an EOM at $\omega_\mathrm{cal}/2\pi = 2.37~\mathrm{GHz}$. The EOM tone is used for phase calibration~\cite{Gorodetsky2010} to calculate the single photon-single phonon optomechanical coupling $g_0/2\pi = 1.3\pm0.3~\mathrm{MHz}$ (at 4.2 K). 

When the temperature is decreased to the millikelvin regime, the mechanical $Q$-factor improves to $Q_\mathrm{m} = 28,800$ as measured using a double-pulse ringdown technique~\cite{Hauer2018}.  We note that the mechanical $Q$-factor decreases with increasing device temperature.  The cooperativity is calculated from the optomechanical interaction rate $\Gamma_\mathrm{om}=4g_0^2 n_\mathrm{cav}/\kappa = 2\pi\times 0.31~\mathrm{MHz}$, where $n_\mathrm{cav} = 230$ is the number of photons in the optical mode. Using these properties, the number of added phonons is calculated to be $n_\mathrm{add} =(C^{-1} + \kappa / 4 \omega_m)^2) / ( 1 + (\kappa / 4 \omega_m)^2)$, where $C=\{n_\mathrm{th}/\mathcal{C},1/\mathcal{C}_\mathrm{qu}\}$ for the ambient and combined baths respectively \cite{Hill2012,Andrews2014}.

\end{document}